# Universal Catalyst Design Framework for Electrochemical Hydrogen Peroxide Synthesis Facilitated by Local Atomic Environment Descriptors

Zhijian Liu[1], Yan Liu[1], Bingqian Zhang[1], Yuqi Zhang[2], Tianxiang Gao[1], Mingzhe Li[1], Xue Jia[3], Di Zhang[3], Heng Liu[3], Xuqiang Shao[2], Li Wei[4,*], Hao Li[3,*], and Weijie Yang[1,*]

[1] School of Energy and Power Engineering, North China Electric Power University, Baoding 071003, P. R. China

[2] Department of Computer Science, North China Electric Power University, Baoding 071003, P. R. China

[3] Advanced Institute for Materials Research (WPI-AIMR), Tohoku University, Sendai, 980–8577 Japan

[4] School of Chemical and Biomolecule Engineering, The University of Sydney, Sydney, New South Wales, Australia

Correspondence and requests for materials should be addressed to W. Y. (yangwj@ncepu.edu.cn) and H. L. (li.hao.b8@tohoku.ac.jp), L. W (l.wei@sydney.edu.au).

# Abstract

Developing a universal and precise design framework is crucial to search high-performance catalysts, but it remains a giant challenge due to the diverse structures and sites across various types of catalysts. To address this challenge, herein, we developed a novel framework by the refined local atomic environment descriptors (*i.e.*, weighted Atomic Center Symmetry Function, wACSF) combined with machine learning (ML), microkinetic modeling, and computational high-throughput screening. This framework is successfully integrated into the Digital Catalysis Database (*DigCat*), enabling efficient screening for 2e$^-$ water oxidation reaction (2e$^-$ WOR) catalysts across four material categories (*i.e.*, metal alloys, metal oxides and perovskites, and single-atom catalysts) within a ML model. The proposed wACSF descriptors integrating both geometric and chemical features are proven effective in predicting the adsorption free energies with ML. Excitingly, based on the wACSF descriptors, the ML models accurately predict the adsorption free energies of hydroxyl ($\Delta G_{OH*}$) and oxygen ($\Delta G_{O*}$) for such a wide range of catalysts, achieving R² values of 0.84 and 0.91, respectively. Through density functional theory calculations and microkinetic modeling, a universal 2e$^-$ WOR microkinetic volcano model was derived with excellent agreement with experimental observations reported to date, which was further used to rapidly screen high-performance catalysts with the input of ML-predicted $\Delta G_{OH*}$. Most importantly, this universal framework can significantly improve the efficiency of catalyst design by considering multiple types of materials at the same time, which can dramatically accelerate the screening of high-performance catalysts. This framework is general and proven effective for predicting $\Delta G_{OH*}$ and $\Delta G_{O*}$, which is expected to be also applicable for the materials design of various catalysis.

# Introduction

The sustainable electrochemical approach to hydrogen peroxide ($H_2O_2$) synthesis is regarded as a cost-effective way for $H_2O_2$ production, presenting an environmentally friendly and energy-efficient method for harnessing renewable energy sources.[1,2] This method also provides a robust energy storage solution by enabling the conversion of electrical energy into chemical energy in the form of $H_2O_2$, thereby bridging energy generation with environmental sustainability.[3]

Currently, the electrochemical synthesis of $H_2O_2$ can be achieved *via* two main processes: the water oxidation reaction (WOR) with water serving as a sole reactant, and the oxygen reduction reaction (ORR) involving both oxygen and water as reactants. Both processes demonstrate promising efficiency under ambient conditions when powered by cost-effective energy sources such as solar or wind energy.[4] These characteristics make them suitable for decentralized production and various applications.[5-7] Among these, the catalytic 2e⁻ WOR has emerged as a more economically viable route for $H_2O_2$ production – by solely relying on water as the feedstock, this process can eliminate the reliance on gaseous reactants. Furthermore, it offers the additional benefit of generating clean hydrogen ($H_2$) at the counter electrode, enhancing both the efficiency and sustainability of the process.[8-10] So far, metal oxides are the predominant anode materials for 2e⁻ WOR in $H_2O_2$ production because some of them have shown excellent stability under oxidizing and alkaline conditions.[11] Various metal oxides, such as $TiO_2$,[12] $SnO_2$,[13] $Sb_2O_3$,[14] $ZnO$,[15] $BiVO_4$,[16] $CaSnO_3$,[17] $CuWO_4$,[18] and $LaAlO_3$,[19] have been reported for $H_2O_2$ generation *via* 2e⁻ WOR. However, these metal oxides still exhibit relatively low intrinsic activity and conductivity; some of them are associated with relatively large overpotentials (*i.e.*, >1.0 V) to overcome the energy barriers, which in turn results in relatively low $H_2O_2$ yields (typically <5 μmol min$^{-1}$ cm$^{-2}$).[20-22] A large-scaling search for promising 2e⁻ WOR catalysts is as pressing as ever.

Although some progress has been achieved in understanding 2e⁻ WOR, most of them predominantly focused on elucidating the relationship between the theoretical overpotential and adsorption free energy of hydroxyl ($\Delta G_{OH^*}$) from a thermodynamic perspective using a theoretical "limiting-potential model".[23,24] However, only analyzing thermodynamics for high-electrode-potential electrocatalysis may sometimes lead to a large discrepancy between theory and experimental observations. For example, back to 2007, study by Vassilev and Koper[26] on the ORR performance has already unrevealed that density functional theory (DFT) based thermodynamic analysis alone cannot fully explain experimental observations. Besides, overpotential is often difficult to be well-defined in experiments, making it hard to perform a direct benchmarking analysis between theory and experiments to validate the model accuracy. To overcome these challenges, microkinetic modeling, which reveals the intricate mechanisms of chemical reactions occurring on the catalyst surface, has emerged as a powerful method to help enhance the understanding of each step within the reaction process. By considering the essential information of both kinetics and thermodynamics, microkinetic modeling can more precisely elucidate the specific reaction pathways, identify key intermediates, and clarify the relationship between reactant bonding strengths and the catalyst's performance, which can in turn predict the key electrochemical indicators (*e.g.*, current density, turnover frequency, Tafel slope) that can be well benchmarked with experimental measurements.[27] Nørskov and colleagues[28] developed the microkinetic volcano model (*i.e.*, the volcano activity model predicting the current density as a function of electrode potential and binding energies of key reaction intermediates) by considering the essential kinetics of ORR, leading to excellent agreement with experimental observations on transition metals in terms of current density at the potentials of interest. This model leads to much higher accuracy than the classic "limiting-potential volcano" for ORR developed two decades ago.[29] Similarly, microkinetic volcano model for 4e$^-$ WOR (*i.e.*, the oxygen evolution reaction, OER) was also developed, which has offered profound insights into the electrochemical OER and filled in many knowledge gaps that the classic OER overpotential model[30] cannot explain. Unfortunately, to the best of our knowledge, understanding 2e$^-$ WOR was still mainly rely on the limiting-potential model[31,32] – a more precise microkinetic model for the 2e$^-$ WOR process has yet to be established.

Furthermore, even a precise microkinetic model is developed, screening promising catalyst candidates still heavy rely on DFT to directly compute the reaction descriptors (*e.g.*, adsorption free energies of key reaction intermediates on a catalyst surface), which is time-consuming and computationally expensive. To meet this challenge, ML offers an effective alternative to accelerate this process.[33,34] In particular, ML models based on Atomic Center Symmetry Function (ACSF) descriptors have shown exceptional predictive accuracy for both energies and forces, exhibiting low prediction errors in many cases.[35-38] By encoding universal local atomic environment features, such as atomic positions and elemental types, these models can reliably predict atomic energies and forces across a wide range of molecular and crystalline materials. Furthermore, the models have demonstrated high accuracy in predicting fundamental material properties, making them suitable for applications involving a wide range of material systems.[37,38] Nonetheless, even focusing on a single type of catalyst, current ML approaches still necessitate complex and time-consuming feature engineering processes, which reduces their scalability and efficiency for broader catalyst screening. Therefore, an automated and universal design framework is urgently needed to be developed for effectively screening various catalyst catagories. The development of such framework may revolutionize the field by significantly improving the screening scale, speed, and accuracy, thereby overcoming the inherent limitations of conventional DFT-based and ML methods.

Motivated by the current limitations in catalyst design, herein, we propose a novel approach by developing and integrating a new type of weighted ACSF (wACSF) descriptors into ML to effectively screen active electrocatalysts, using the less-explored 2e$^-$ WOR process as a typical example. Based on the developed wACSF descriptors, XGBoost regression (XGBR) models were developed, which are capable of accurately predicting the adsorption free energies of hydroxyl ($\Delta G_{OH*}$) and oxygen ($\Delta G_{O*}$) across a wide range of materials at the same time, including metal alloys, metal oxides and perovskites, and single atom catalysts (SACs), achieving high predictive accuracies ($R^2$ = 0.84 and 0.91, and RMSE = 0.52 and 0.65 eV, respectively for $\Delta G_{OH*}$ and $\Delta G_{O*}$) at their optimal cutoff radii. The wACSF descriptors, integrating central and coordination geometric features, can be applied across different catalyst types, addressing the transferability limitations of conventional ML-assisted catalyst design. Our well-developed automatic extraction workflow of descriptors from material structures significantly enhances the screening efficiency. Additionally, by integrating DFT and microkinetic analysis, we constructed a precise microkinetic volcano model for 2e$^-$ WOR, providing a powerful tool to rapidly identify high-performance 2e$^-$ WOR electrocatalysts with the input of ML-predicted adsorption free energies. This combined strategy not only conserves computational resources but also accelerates the precise design of catalysts, demonstrating its potential for large-scale material discovery.

# Results

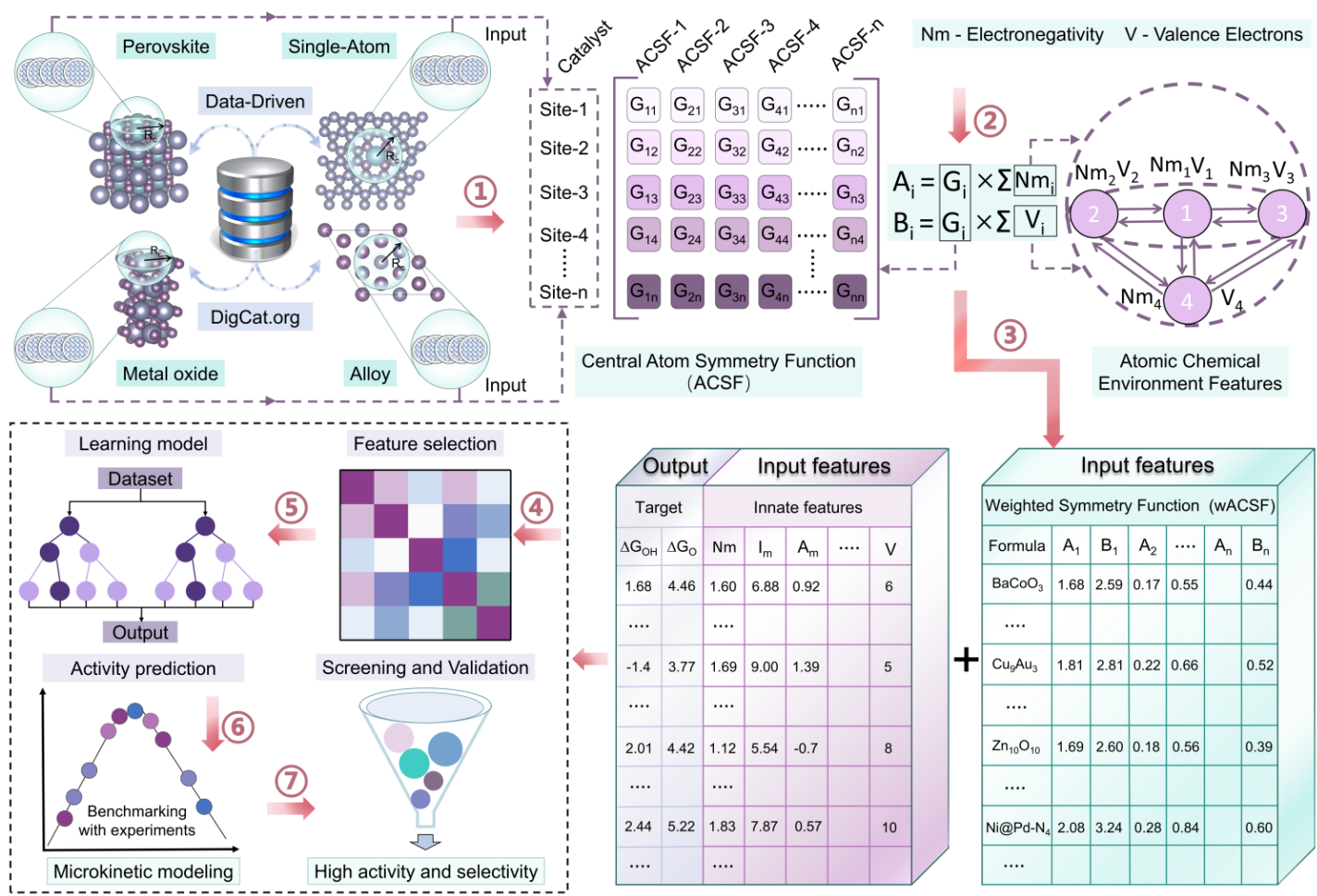

**Figure 1 | Overview of the catalyst design framework developed in this study.** This framework comprises (1) feature extraction from the Digital Catalysis Database (*DigCat*), (2) feature calculation, (3) data generation, (4) feature selection *via* Pearson correlation and recursive elimination, (5) ML training and testing, (6) microkinetic modeling and benchmarking analysis with available experimental data from the *DigCat* database, and (7) ML-accelerated catalyst screening.

**Overview of the Universal Framework:** The proposed framework is outlined in **Figure 1**, adhering to the steps detailed therein. First, four typical types of materials are selected (**Figure 1, Step 1**), including metal alloys, metal oxides and perovskites, and SACs, from the recently developed *DigCat* database (DigCat.org). Afterward, the radial and angular symmetry function values ($G_i$) of active sites on various catalysts were calculated under different combinations of symmetry function parameters (**Supplementary Note 1**).[39] Subsequently, the chemical environment features of the atoms were constructed by summing the electronegativity (Nm) and valence electron count (V) of the central and coordinating atoms of the catalysts.[40] The calculated $G_i$ were then multiplied by the chemical environment features ($\Sigma Nm_i$ and $\Sigma V_i$) (**Figure 1**, **Step 2**). This process involves differentiating the contributions of $G_i$ for each catalyst through weighting based on electronegativity and valence electron count, resulting in the "wACSF" descriptors (**Supplementary Note 2**). The two types of weighted symmetry function values ($A_i$ and $B_i$) were merged with the intrinsic feature values of the catalysts and used as the independent variables (**Figure 1, Step 3**). The adsorption free energy data of $\Delta G_{OH*}$ and $\Delta G_{O*}$ served as the target values (*i.e.*, the outputs of ML), establishing a connection between the adsorption characteristics at the catalytic sites and their numerical fingerprints. This process is fully automated through the implementation of our code. Subsequently, feature analysis and recursive elimination are conducted to discard redundant features (**Figure 1**, **Step 4**). Advanced ML algorithms are then employed to develop the effective model (**Figure 1**, **Step 5**). The optimal ML model will be used to predict the adsorption free energies of new catalysts. These adsorption free energies will be incorporated into a microkinetic volcano model, which will be benchmarked against experimental data to evaluate the predictive accuracy of the microkinetic model (**Figure 1**, **Step 6**). Finally, potential high-performance catalysts will be rapidly identified by the microkinetic model through a screening process with the input of ML-predicted adsorption free energies (**Figure 1**, **Step 7**).

**Database and feature generation:** We have assembled an adsorption free energy dataset with 962 materials from the *DigCat* database, comprising 426 intermetallic alloys, 160 metal oxides (excluding perovskites), 250 perovskites, and 126 SACs (**Figure 1, Step 1**; **Supplementary Dataset 1**). These datasets were originally extracted from published literature, encompassing a variety of surfaces, adsorption sites, and values of adsorption free energies (*i.e.*, $\Delta G_{OH*}$ and $\Delta G_{O*}$). In terms of the features as the independent variables to describe these 962 materials, based on the primary ACSF features, candidate features can be created by weighting the symmetry functions with the electronegativity and valence electron count of the coordination environment (**Figure 1**, **Step 2**). These features are then automatically compiled and output into a table. Automated feature extraction is a crucial component of the training and prediction process (**Supplementary Note 3**). The structural information of each material's central atom, including coordination element information, establishes a unique channel. The consistent characterization of active sites and their surrounding environments across various materials ensures the broad applicability. In addition to the wACSF descriptors, elemental intrinsic properties (*i.e.*, electronegativity, valence electrons, first ionization energy, electron affinity, atomic number, and atomic radius) were also included as potential features. Ultimately, each dataset was standardized to include 134 descriptors. These 134 features were then applied to the adsorption free energy dataset, resulting in a 962×134 feature matrix with two endpoint vectors (**Figure 1**, **Step 3**), $\Delta G_{OH^*}$ and $\Delta G_{O^*}$, which would be used in the subsequent feature engineering. A key advantage of this feature design strategy is that these descriptors can be applicable to various materials consisting of different elements.

To enable rapid automated extraction of wACSF values and elemental features for the 962 catalysts' active sites and their local environments, we developed a Python code (refer to the **Code Availability Section**), which can calculate the wACSF values for different types of materials, output their intrinsic features, and generate a feature set to effectively obtain numerous descriptors for target structures. The symbols and definitions of all the selected descriptors are detailed in **Supplementary Table 1**. Our method automatically generates symmetry function features for active sites in various materials by using CIF files as input and reading the catalyst active site information provided by the user, thereby automating the generation of these features. The automated extraction model retrieves intrinsic features by parsing element information from catalyst names, extracting features from our library, and integrating them into the training process (**Supplementary Dataset 2**).

**Feature engineering:** Next, we applied Pearson correlation analysis and the recursive feature elimination (RFE) method to identify the critical features in this study (**Figure 1**, **Step 4**),[41,42] isolating the features that substantially enhance model performance (for further details, refer to **Methods Section** in **Supplementary Notes 4** and **5**). Following this feature selection process, we identified 21 key descriptors (including 5 atomic intrinsic properties and 16 wACSF features) for $\Delta G_{OH^*}$ and 19 descriptors (including 3 atomic intrinsic properties and 16 wACSF features) for $\Delta G_{O^*}$ (**Figure 2**; for details of these features, refer to **Supplementary Dataset 3**). These descriptors demonstrated substantial predictive power during preliminary ML training. As shown in **Figure 2**, a deeper color corresponds to a more positive Pearson correlation coefficient between the pair of features in the corresponding row and column. A correlation value of 1 indicates the highest correlation between the two descriptors. A high correlation may result from the intrinsic relationships among parameters within the wACSF, which can cause elevated correlations between the features derived from them. The design of the symmetry functions might lead to features with similar properties. Nevertheless, these features can still offer unique information that can enhance model performance, which is crucial to improve the model's predictive accuracy regarding adsorption free energies. Nonlinear models, such as the XGBR method,[43] are particularly adept at managing multicollinearity among features and extracting valuable information, making these features beneficial. In practical applications, particularly where model performance is paramount, retaining these correlated features allows for the full utilization of their information, thereby enhancing the model's predictive accuracy.

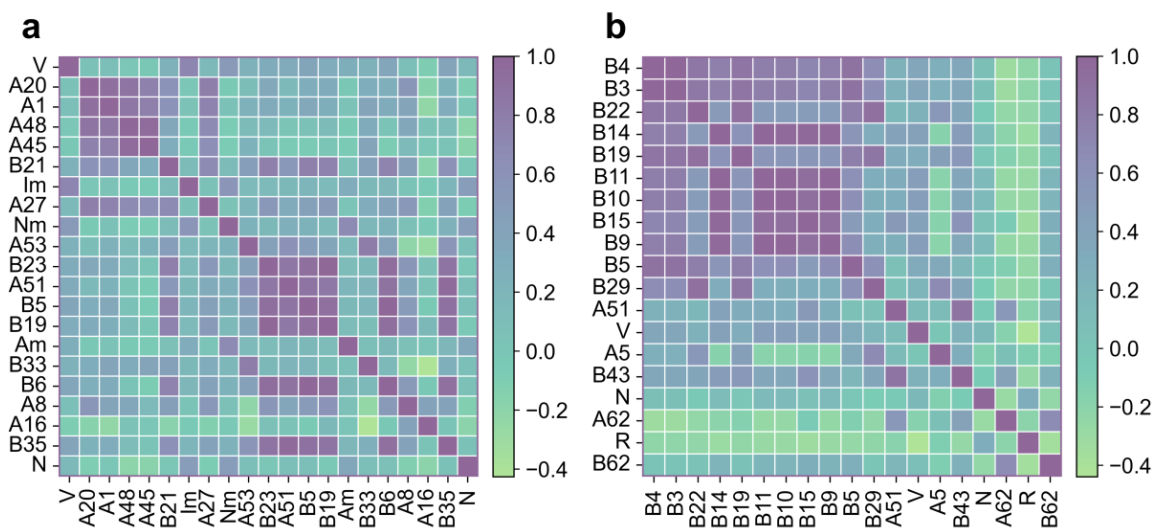

**Figure 2 | Pearson correlation matrix of the remaining features after feature engineering.** (a-b) Heatmap of the Pearson correlation coefficients between descriptors to predict (a) $\Delta G_{OH*}$ and (b) $\Delta G_{O*}$. The values in the grid represent the correlation coefficients between the two corresponding descriptors. A deeper color corresponds to a more positive Pearson correlation coefficient.

**Performance and Validation of ML Models:** For ML modeling (**Figure 1**, **Step 5**), first, the entire dataset of adsorption energies was randomly partitioned into a training set (80%) and a test set (20%) (**Supplementary Note 6**). The training set was utilized for model parameter tuning, while the test set, which remained unseen during training, was used for independent validation. To ensure accurate predictions of adsorption free energies, models trained with various ML algorithms were compared. To determine the optimal hyperparameters for each model, a 10-fold cross-validation was performed on the training set. Models including random forest regression (RFR),[44] kernel ridge regression (KRR),[45] XGBR,[46] ridge regression (RR),[47] k-nearest neighbors (KNN),[48] gradient boosting regression (GBR),[49] support vector regression (SVR),[50] and extra trees regression (ETR)[51] were evaluated (**Supplementary Note 7**). For each model, multiple modeling parameters were tested and evaluated, to reduce the bias in model performance comparison. We selected the most precise and robust ML models based on the coefficient of determination ($R^2$) and root mean square error (RMSE) (**Supplementary Note 8**).[52] **Figure 3** illustrates that models trained with ETR, SVR, KNN, and RR exhibit notably lower predictive performance compared to those trained with RFR, XGBR, GBR, and KRR. Moreover, the predictions from ML models trained with XGBR and GBR closely align with the target values, demonstrating the superior performance of these models. Based on both the training and predictive performance, XGBR is identified as the optimal method for predicting adsorption free energies in our case. Therefore, ML models for $\Delta G_{OH*}$ and $\Delta G_{O*}$ were developed using the XGBR algorithm.[53,54]

To note, the choice of truncation radius ($R_c$) significantly impacts the model's predictive performance. Specifically, we used $R_c$ to define the size of the region surrounding the central atom.[55] The selection of $R_c$ is critical for model accuracy and robustness, as it defines the extent of the local environment considered. Therefore, optimizing the $R_c$ is essential to enhance model performance.[56,57] Herein, different $R_c$ values were tested to evaluate their effects on the performance metrics (*i.e.*, $R^2$ and RMSE) during training and test. Interestingly, results indicate that for $\Delta G_{OH*}$, the optimal $R_c$ is ~7 Å, yielding an $R^2$ of 0.84 and an RMSE of 0.52 eV for the test set (**Supplementary Figure 1**). For $\Delta G_{O*}$, the optimal $R_c$ is ~5 Å, with the test set achieving an $R^2$ of 0.91 and an RMSE of 0.65 eV. These choices are based on the observation that the model performance metrics converge when $R_c$ exceeds 7 or 5 Å, suggesting that adsorption free energy information is predominantly determined within the local environment. Consistent with the findings reported in Ref. 60, which noted an improved model accuracy when transitioning from global to local descriptors due to the strong dependence of adsorption strength on the local environment, a moderate $R_c$ value can save computational resources and training time, and meanwhile, maintain a high-level accuracy in ML prediction. This is because an unnecessarily large $R_c$ may increase computational complexity and the difficulty in ML training. To note, another interesting phenomenon is that $\Delta G_{OH*}$ requires a larger $R_c$ than $\Delta G_{O*}$, which is plausibly because the bonding strength of an adsorbate with a high electron affinity (*e.g.*, radical adsorbates like OH) is generally influenced by longer-range interactions.[58]

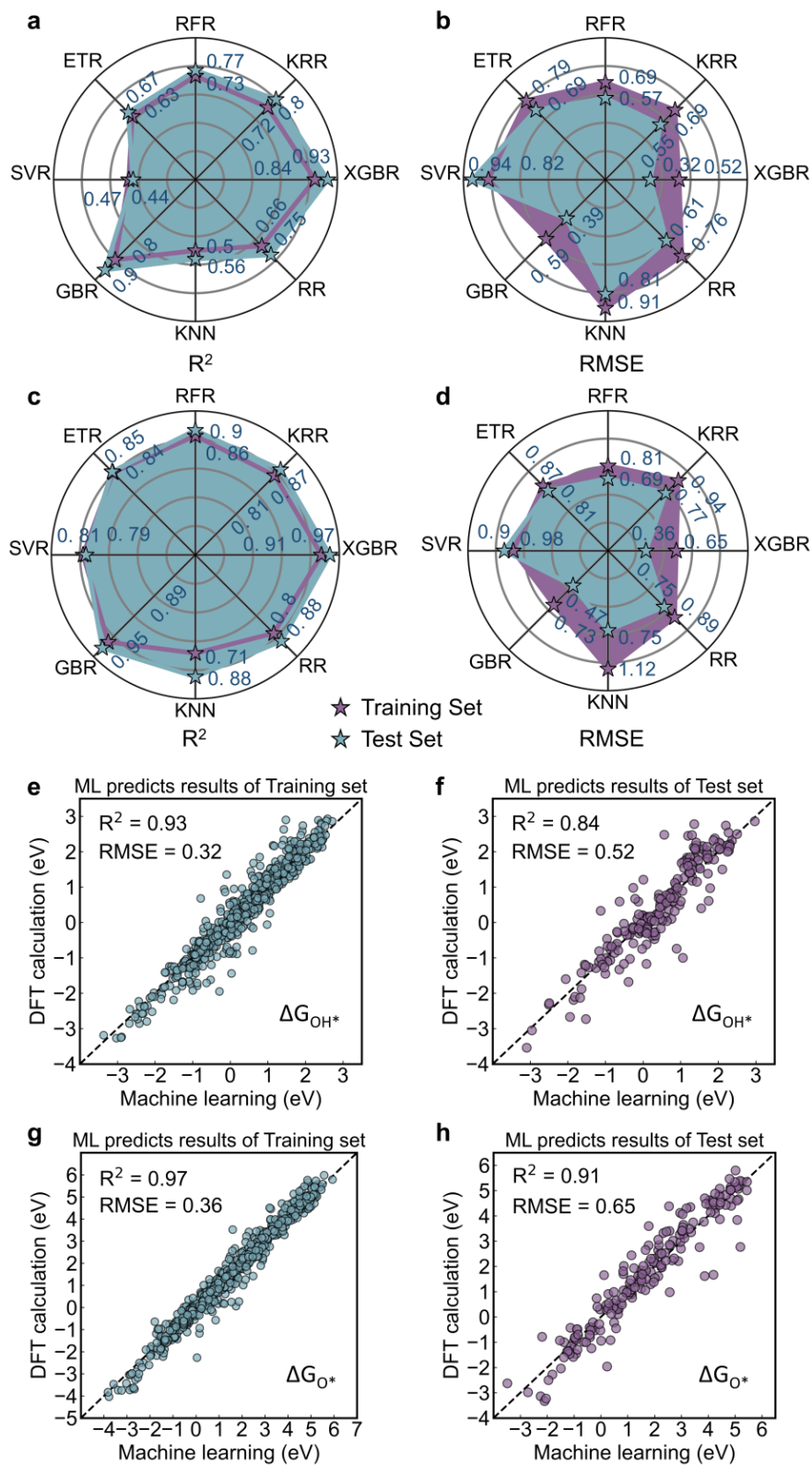

**Figure 3 | Evaluation of ML regression performance for adsorption free energy predictions.** Panels (a-b) display the (a) R² and (b) RMSE for the prediction of ΔG$_{OH*}$. Panels (c-d) display the (c) R² and (d) RMSE for the prediction of ΔG$_{O*}$. Panels (e-f) display the comparison between the ΔG$_{OH*}$ calculated by DFT (ΔG$_{OH*}$, DFT)

and predicted by the ML model ($\Delta G_{OH*}$, ML), respectively in the (e) training and (f) test sets. Panels (g-h) display the comparison between the $\Delta G_{O*}$ calculated by DFT ($\Delta G_{O*}$, DFT) and $\Delta G_{O*}$ predicted by the ML model ($\Delta G_{O*}$, ML), respectively in the (g) training and (h) test sets.

**Figure 3e-h** shows the comparison between the ML-predicted adsorption free energies from the optimal XGBR model and those calculated by DFT. For $\Delta G_{OH*}$, the R² between DFT and ML predictions are 0.93 and 0.84, respectively in the training and test sets. The RMSE, used as the loss function to evaluate optimization errors, is 0.32 eV for the training set and 0.52 eV for the test set (**Figure 3e-f**). For $\Delta G_{O*}$, the R² values are 0.97 for the training set and 0.91 for the test set. The corresponding RMSE values are 0.36 eV for the training set and 0.65 eV for the test set (**Figure 3g-h**). The above results clearly demonstrate that these ML models can provide high accuracies in predicting adsorption free energies.

To further show the high accuracy of our method, **Table 1** compares our models' performance with the previously reported methods for predicting adsorption energies. In comparison to previous ML models using Surface Center-Environment descriptors,[59] which achieved RMSEs of 0.54 ($\Delta G_{OH*}$) and 0.96 eV ($\Delta G_{O*}$), our model, which utilizes simpler descriptors, shows lower errors in the test sets. Although our model's RMSE in the test set is slightly higher compared to the lowest RMSE achieved with intrinsic and electronic/geometric descriptors in existing studies (**Table 1**),[60] it provides a significant advantage by offering universal descriptors applicable across various materials in one ML model, which is not limited to the type of material in one training-and-testing process. Our model not only overcomes the limitations of conventional methods but also offers a more flexible tool to screen various types of materials at the same time, through a simplified and automated process. Furthermore, the accuracy of our model is comparable to models using Three-Dimension Coordinates (RMSE: 0.63 eV) and Cavity Features (RMSE: 0.56 eV) descriptors (**Table 1**),[61,62] which require costly DFT computations and complex feature interpretations. In contrast, our proposed descriptors are more accessible and practical, providing a cost-effective alternative.

To summarize, ML models with the developed universal descriptors have been developed, capable of effectively predicting the adsorption free energies for various types of materials in one ML model. This method demonstrates notable practical advantages for large-scale and diverse material screening tasks.

Table 1 | Comparison of the predictive performance among different studies.

| Descriptor | Target | Catalyst | Model | $R^{2[a]}$ | RMSE [a] (eV) | Ref. |
|---|---|---|---|---|---|---|
| Weighted ACSF | $\Delta G_{OH*}$ | Four Catalysts | XGBR | 0.84 | 0.52 | This Work |
| Weighted ACSF | $\Delta G_{O*}$ | Four Catalysts | XGBR | 0.91 | 0.65 | This Work |
| Surface Center-Environment | $\Delta G_{OH*}$ | $ABO_3$ | RFR | 0.79 | 0.54 | 59 |
| Surface Center-Environment | $\Delta G_{O*}$ | $ABO_3$ | RFR | 0.80 | 0.93 | 59 |
| Properties of Metals and Alloys | $\Delta G_{O*}$ | Binary Alloys | NN | - | 0.34 | 60 |
| Inherent Features | $E_{ads,OH}$ | 2D Materials | XGBR | 0.85 | 0.18 | 53 |
| Inherent Features | $E_{ads,H}$ | 2D Materials | XGBR | 0.89 | 0.1 | 53 |
| Three-Dimension Coordinates | $\Delta G_{OOH*}$ | SACs | MLP | - | 0.63 | 61 |
| Cavity Features | $E_{ads}$ | SACs Doped Carbons | RFR | 0.93 | 0.56 | 62 |
| Inherent Features | $\Delta G_{OH*}$ | SACs | XGBR | 0.96 | 0.29 | 63 |
| Inherent Features | $\Delta G_{H*}$ | SACs | XGBR | 0.91 | 0.27 | 63 |
| Electronic and Geometric | Adsorption Enthalpy | Metal Oxides | SISSO | - | 0.18 | 64 |

[a] Note: the $R^2$ and RMSE are the performance indicators of the test sets.

**Importance of Descriptors and Their Physicochemical Interpretations:** To understand the impact of these features on the predicted adsorption free energies, we analyzed the importance of each feature and its contribution to predicting $\Delta G_{OH*}$ and $\Delta G_{O*}$, represented by the SHAP (SHapley Additive exPlanations) values (**Figure 4**). SHAP is a powerful analytical method to interpret ML results based on the Shapley values from game theory (**Supplementary Note 9**).[65] In **Figure 4b** and **4d**, the *x*-axis represents SHAP values, where a more positive SHAP value indicates a more positive contribution of the feature to the adsorption free energy. Each solid circle corresponds to the SHAP value assigned to a specific feature affecting the catalysts' adsorption free energies. The color gradient, ranging from cyan to purple, indicates the relative magnitude of the feature values, with cyan representing higher values and purple representing lower values.

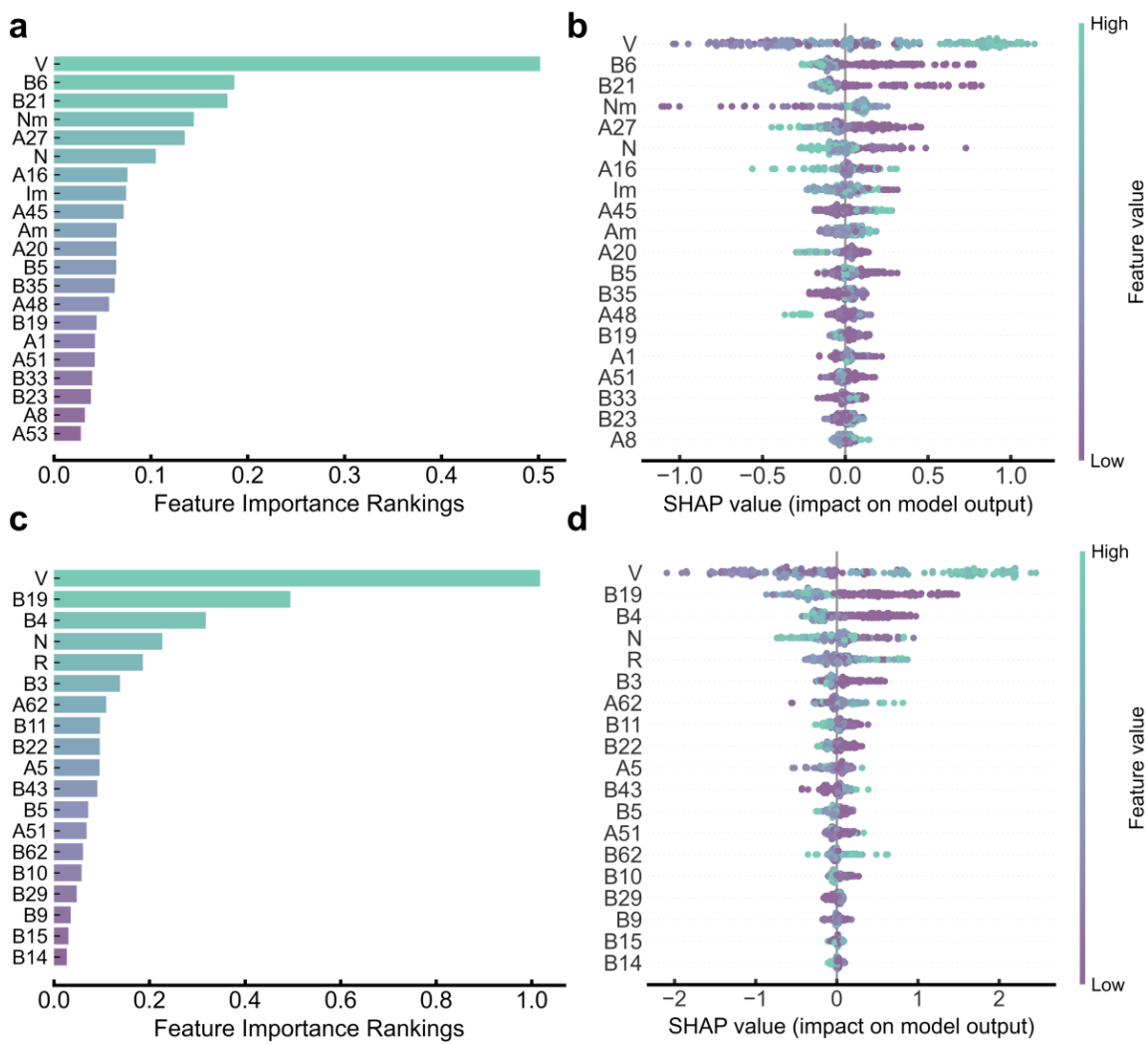

**Figure 4 | SHAP analysis of feature importance for various materials based on the XGBR model.** (a,c) Feature importance based on average SHAP values for the (a) $\Delta G_{OH*}$ model and (c) the $\Delta G_{O*}$ model. (b,d) SHAP sensitivity analysis of individual features for the (b) $\Delta G_{OH*}$ model and (d) the $\Delta G_{O*}$ model, illustrating the impact of each feature on the model's output.

A wider distribution of SHAP values indicates a greater sensitivity of the corresponding feature to the target adsorption free energy. For $\Delta G_{OH*}$, the three descriptors with the widest distribution of SHAP values are Nm (central atom electronegativity; importance score: 0.50), B6 (radial symmetry function weighted by coordination valence electrons; importance score: 0.19), and B21 (angular symmetry function weighted by coordination valence electrons; importance score: 0.18). For $\Delta G_{O*}$, the three descriptors with the widest distribution of SHAP values are Nm (central atom electronegativity; importance score: 1.02), B19 (angular symmetry function weighted by coordination valence electrons; importance score: 0.50), and B4 (radial symmetry function weighted by coordination valence electrons; importance score: 0.32). These descriptors exhibit high feature importance scores. In contrast, the majority of data points for other features are clustered around a SHAP value of 0, indicating lower sensitivity and minimal impact on the adsorption free energy. The findings from the SHAP analysis, along with the identified key descriptors, provide insights into their physicochemical significance, which can be effectively explained through theoretical evaluation. The central atom's Nm shows a high sensitivity to the output, as indicated by its broad SHAP value distribution and high importance. Electronegativity reflects an atom's ability to donate or accept electrons - atoms with higher electronegativity tend to accept electrons, while those with lower electronegativity are more prone to lose electrons. The color distribution reveals that lower V values (purple) negatively impact $\Delta G_{OH*}$ and $\Delta G_{O*}$, while higher V values (cyan) have a positive impact. This implies that higher V corresponds to higher output values (**Figure 4b** and **4d**). The fundamental physicochemical principle is that electronegativity affects the electronic transfer and interaction strength between the central atom and the adsorbate. Atoms with higher electronegativity are more likely to attract electrons. Therefore, accurately describing materials' performance in ML modeling requires considering the inherent properties of the central metal atom.[66] The above results indicate that $\Delta G_{OH*}$ and $\Delta G_{O*}$ are highly correlated with the electron-donating and -accepting properties of the atoms in the materials. This finding aligns with previous research by Zhu et al.,[67] which demonstrates that differences in electronegativity between elements can lead to significant charge redistribution, thereby modulating adsorption affinity.

Similarly, B6, B4, B21, and B19 also exhibit high sensitivity to the adsorption free energies due to their broad distribution (**Figure 4b** and **4d**). B6 and B4 are radial symmetry function descriptors weighted by coordination valence electrons, while B21 and B19 are angular symmetry function descriptors weighted by coordination valence electrons. The color distribution shows that lower values of B6, B4, B21, and B19 (purple) positively impact $\Delta G_{OH*}$ and $\Delta G_{O*}$, generally aiding in predicting those catalysts with stronger adsorption capacity. This may be because lower values of these descriptors indicate a smaller product of the coordination atom's valence electron count and the central atom's symmetry function, which could imply a more uniform electronic environment around the central atom, facilitating the reactant adsorption and stabilizing reaction intermediates. This might also suggest weaker interactions between the coordination atoms and the central atom, potentially stabilizing the adsorbate species.

The fundamental physicochemical principle of wACSF is that radial (or angular) symmetry function descriptors consider the radial (or angular) distribution of coordination atoms around the central atom, where the weighted treatment by coordination valence electrons further incorporates the electronic contribution of coordination atoms. Therefore, B6 and B21 for $\Delta G_{OH*}$ (and B4 and B19 for $\Delta G_{O*}$) reflect the interaction strengths and electronic distribution between the central atom and its coordination atoms in the radial and angular directions. This feature provides essential information for describing complex angular-dependent interactions during adsorption and is crucial for modeling the local chemical environment of atoms.[42,55]

In summary, the top three important descriptors (*i.e.*, electronegativity, radial distribution, angular distribution) comprehensively reflect the local chemical environment and electronic distribution between the central atom and coordination atoms from different physicochemical perspectives. Two of these key features are new descriptors proposed by this study, highlighting their role as effective universal descriptors and their critical importance in predicting adsorption free energies.

**Microkinetic Modeling and Catalyst Screening:** To accurately predict the activities of 2e⁻ WOR catalysts and screen the materials mentioned above, herein, we developed a microkinetic model for 2e⁻ WOR, which predicts the reaction exchange current density ($j_0$) as a function of electrode potential and $\Delta G_{OH*}$ (**Figure 1, Step 6**). The corresponding steps and rate equations are detailed in **Equations 20-37** (refer to the **Methods Section** in **Supplementary Note 10**), with the associated code accessible in our *CatMath* on-the-cloud platform.[68] Typically, a volcano plot can be derived based on the relationship between $\Delta G_{OH*}$ and $j_0$ (**Figure 5a**). This plot serves as a crucial role to evaluate the catalytic activity of 2e⁻ WOR electrocatalysts.

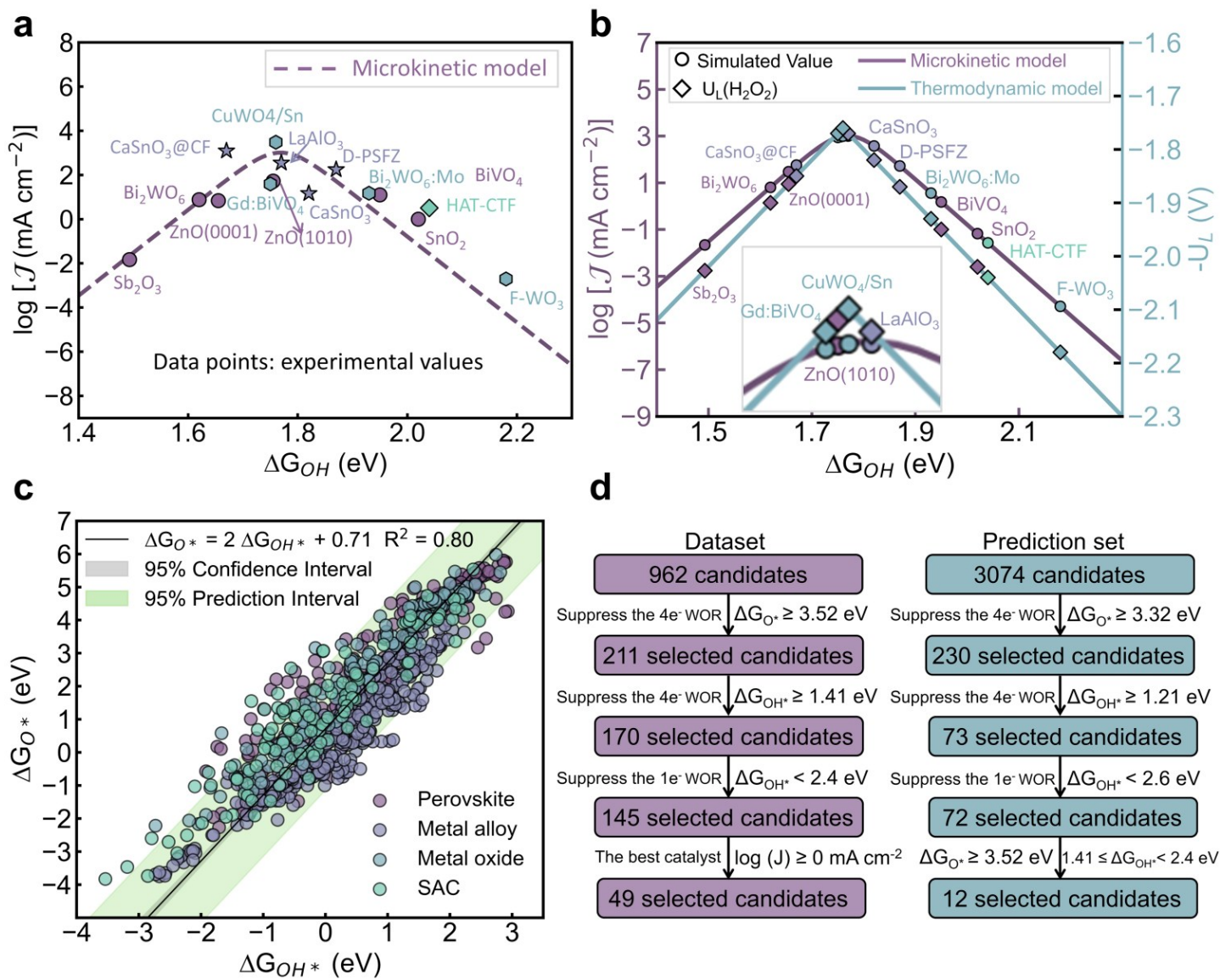

**Figure 5 | Microkinetic and Thermodynamic Volcano Models for 2e⁻ WOR and the Catalyst Screening Process.** (a) Simulated microkinetic volcano model for 2e⁻ WOR at 2.4 V *vs*. RHE. The data points were extracted from experiments (**Supplementary Table 2**; also available in the *DigCat* database). (b) Microkinetic and thermodynamic volcano models plotted with the calculated exchange current density and theoretical limiting-potential values. (c) Scaling relation between $\Delta G_{OH*}$ and $\Delta G_{O*}$ for 962 catalyst surfaces. (d) Workflow of the data screening process, illustrating the number of candidates selected after each screening step.

Based on the reaction mechanisms, we simulated the relationship between the 2e⁻ WOR current density and OH* binding free energy of catalysts, as indicated by the dashed line in **Figure 5a** (the simulation methods can be found in **Supplementary Note 10**). All the models are benchmarked by experimental data from various catalysts (the data sources can be found in the *DigCat* database, which are also available in **Supplementary Table 2**), represented by the data points in **Figure 5a**. Excitingly, it can be clearly seen that simulated microkinetic volcano has excellent agreement with the experimental observations.

In contrast to **Figure 5a**, which employs experimental values, **Figure 5b** presents the current density values calculated based on the microkinetic model. **Figure 5b** depicts a dual-model plot of microkinetic and the (previously reported)[24] thermodynamic volcanoes for metal oxide catalysts. The cyan solid line in **Figure 5b** illustrates the thermodynamic activity volcano determined by the limiting-potential (calculation method are detailed in **Supplementary Note 11**). Interestingly, both the kinetic and thermodynamic volcanos demonstrate similar activity trends. At 2.4 V vs. reversible hydrogen electrode (RHE), the peak of the thermodynamic volcano occurs at $\Delta G_{OH*}$ = 1.76 eV, while the peak of the kinetic volcano is at 1.77 eV. On the left-leg of the thermodynamic volcano, the rate-limiting step is the oxidation of OH* to $H_2O_2$, while on the right-leg, the limiting step is the activation of $H_2O$ to OH*. In contrast, in the microkinetic volcano, these rate-limiting steps are reversed, with the activation of $H_2O$ to OH* becoming the limiting step on the left-leg and the oxidation of OH* to $H_2O_2$ on the right-leg. This suggests that though a simplified limiting-potential model can capture the performance trend of 2e⁻ WOR in experiments, it may fail to accurately predict the reaction mechanism – this indicates that, a comprehensive microkinetic model, which can pass the benchmarking analysis with experimental values (**Figure 5a**), would be more reliable for the further screening of potential 2e⁻ WOR catalysts.

From both the simulated and experimental results, catalysts such as $CaSnO_3$@CF,[72] $CuWO_4$/Sn,[71] $LaAlO_3$/FTO,[20] ZnO/FTO (1010),[15] and Gd-doped $BiVO_4$ (Gd:$BiVO_4$)[16] are near the peak of the microkinetic volcano model (**Figure 5a**). Therefore, they are considered as the best-performing WOR electrocatalysts, which is also confirmed by the available experimental data in the *DigCat* database. In addition to conventional metal oxides, perovskites such as $Pr_{1.0}Sr_{1.0}Fe_{0.75}Zn_{0.25}O_{4-\delta}$ (D-PSFZ), metal-free organic networks like covalent triazine frameworks hexaazatriphenylenes (HAT-CTF), and doped or modified metal oxides also show promising performance. These further validate the microkinetic model's accuracy and applicability, demonstrating exceptionally good consistency with experimental results across various types of materials.

Furthermore, the selectivity of catalysts is another crucial aspect that should be considered. In the screening process, we considered the catalysts' selectivity as a key criterion to assess the potential of various catalysts for 2e⁻ WOR. We derived quantitative criteria based on the reaction's thermodynamics. To achieve target selectivity, the free energy of OH* on the catalyst surface should be more negative than that of OH·(aq) (*i.e.*, $\Delta G_{OH*}$ <2.4 eV) to prevent the formation of OH· radicals. This sets an upper limit for the free energy of $H_2O_2$ on the surface. In other words, to prevent single-electron oxidation (**Equations 38-40** in **Supplementary Note 11**) and to achieve two-electron oxidation (**Equations 41-43** in **Supplementary Note 11**), it is required that $\Delta G_{OH*}$ should be more negative than 2.4 eV. Additionally, since the formation energy of $H_2O_2$ in solution ($\Delta G_{H2O2}$) is constant at 3.52 eV, catalysts with a weaker O* adsorption strength ($\Delta G_{O*} \geq 3.52$ eV) are favorable for $H_2O_2$ formation (*i.e.*, 2e⁻ WOR), while those with values exceeding 3.52 eV tend to favor the OER pathways (*i.e.*, 4e⁻ WOR).

Previous research has established that the adsorption free energies of various reaction intermediates (O*, OH*, and OOH*) are suitable descriptors for the activity and selectivity of WOR pathways.[24] For effective 2e⁻ WOR catalysts, the adsorption energy of OH* should be optimized. A proportional relationship between the adsorption free energies of two key intermediates was established to describe the adsorption behavior of 962 materials within the *DigCat* database. **Figure 5c** presents the scaling relationship between $\Delta G_{OH*}$ and $\Delta G_{O*}$ ($\Delta G_{O*}$ = 2$\Delta G_{OH*}$ + 0.71). The slope of the proportional line for the catalyst was found to be ~2, aligning with the previously reports for metal oxide catalysts.[69,70] Considering the proportional relationship between O* and OH*, this sets a lower limit for the free energy of OH* to approximately $\Delta G_{OH*} \geq (3.52/2 - 0.71/2)$ eV (*i.e.*, ~1.41 eV).

In summary, considering the binding free energies of OH* and O*, systems with good selectivity for generating $H_2O_2$ through 2e⁻ WOR should meet the following criteria: $\Delta G_{O*} \geq 3.52$ eV and 1.41 eV $\leq \Delta G_{OH*}$ < 2.4 eV. These criteria were applied to the materials in the *DigCat* database and the prediction set for preliminary screening.

**Selection and Validation of Highly Active and Selective 2e⁻ WOR Catalysts:** The workflow for the screening process is exemplified in **Figure 5d** (also overviewed in **Figure 1, Step 7**). Initially, a preliminary screening was conducted on 962 catalysts from the *DigCat* database, resulting in the selection of 145 candidates that met the criteria (**Supplementary Dataset 4**). These catalysts demonstrated OH* adsorption energies within an optimal range of 1.41 to 2.4 eV, with $\Delta G_{OH*}$ values consistent with the conditions required for high selectivity in $H_2O_2$ production. After applying a more stringent criterion, specifically log(J) ≥0 mA cm$^{-2}$, the candidates were shortlisted to 49 (**Supplementary Dataset 4**). These 145 catalysts were then subjected to further microkinetic modeling analysis (**Figure 6a**). After additional literature search, catalysts marked with stars in the figure, such as $LaAlO_3$, $Zn_{10}O_{10}$, and $Ti_6O_{12}$, have also been synthesized in experiments and demonstrated excellent 2e$^-$ WOR activities.[19,15,11] This suggests a high likelihood of successful experimental perspective for these ML-predicted catalysts, demonstrating that our predictive model and the screening process are reliable and effective in identifying efficient catalysts with promising practical applications.

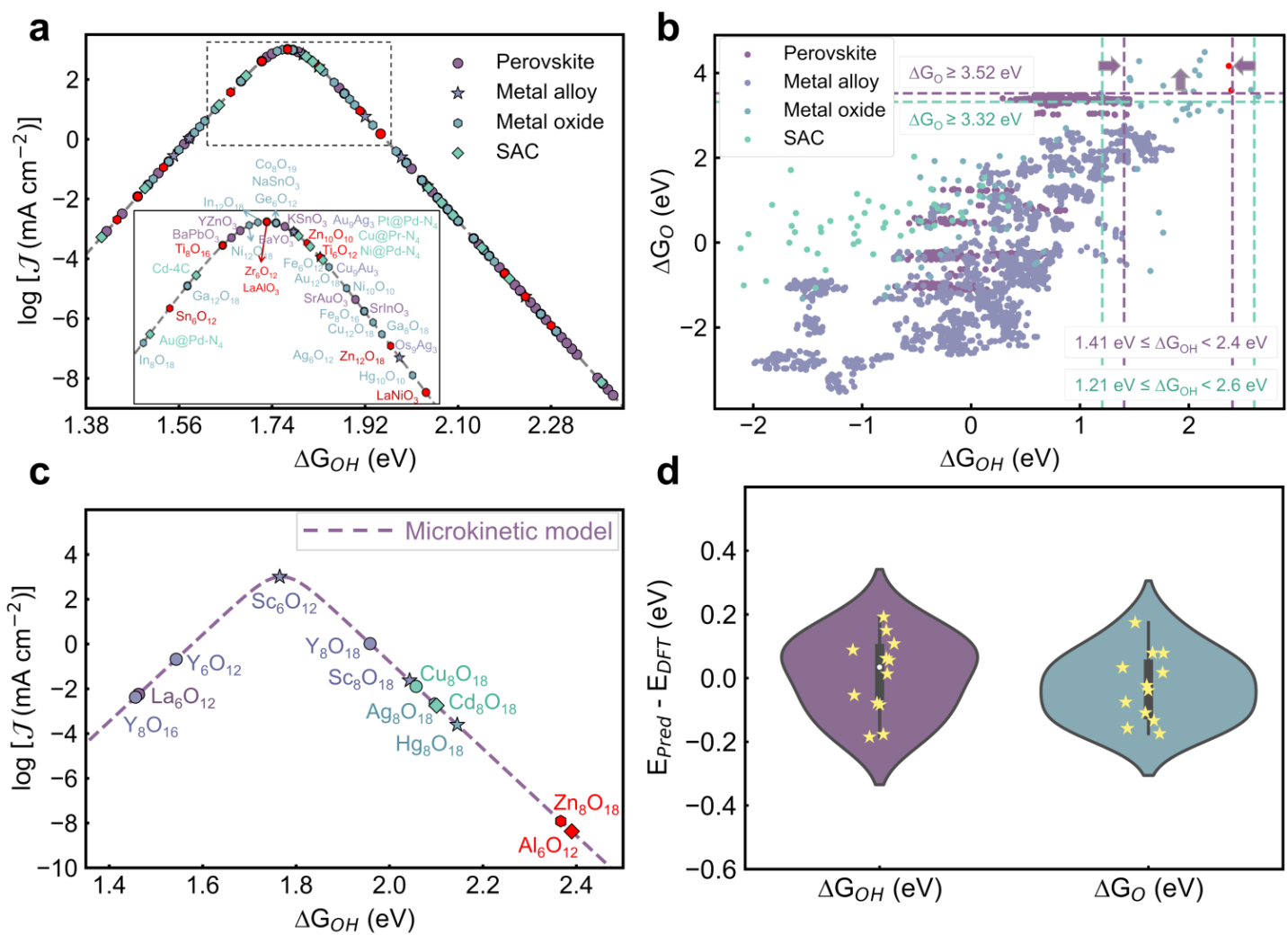

**Figure 6 | Performance Analysis and Validation of the Selected Catalysts.** (a) Microkinetic volcano model predicts the 2e$^-$ WOR performance of 145 materials from the preliminary screening process in the *DigCat* database.

(b) Selectivity plot for the ML-predicted adsorption free energies of 3074 materials. (c) Microkinetic volcano model plotted with the 12 materials selected from the prediction from (b). (d) Violin plot showing the distribution of differences between the predicted and DFT-calculated adsorption free energies for the 12 materials selected from (c).

To perform a larger-scale catalyst screening, a total of 3,074 new structures were created across four categories of catalysts: 2,381 metal alloys, 74 metal oxides (excluding perovskites), 569 perovskites, and 50 SACs, to predict $\Delta G_{OH*}$ and $\Delta G_{O*}$ (**Supplementary Dataset 4**). Based on this data, screening criteria and microkinetic volcano models will be further applied to evaluate the activity and selectivity of both existing database materials and newly generated materials, identifying the most efficient catalysts for 2e⁻ WOR. Similarly, a preliminary screening was performed on these 3074 materials in the prediction set to validate the selectivity and activity screening criteria. **Figure 6b** depicts the adsorption free energies predicted by the ML model developed above. To enhance screening accuracy, a prediction error margin of 0.20 eV was considered. This adjustment aims to minimize uncertainty in model predictions and identify candidates with greater practical potential. The screening criteria were set as 1.21 eV ≤ $\Delta G_{OH*}$ < 2.6 eV and $\Delta G_{O*}$ ≥ 3.32 eV, leading to the finalist with 72 suitable catalysts (**Supplementary Dataset 4**), including metal oxides and perovskites, and metal alloys. These candidates have potential for $H_2O_2$ production in the 2e⁻ WOR process. Notably, $Zn_8O_{18}$[73] and $Al_2O_3$[74] have been experimentally confirmed to exhibit 2e⁻ WOR catalytic activity. This demonstrates that our screening criteria effectively identify materials with practical catalytic potential and provides valuable references for future catalyst development and optimization.

Subsequently, more stringent screening criteria were applied (1.41 eV ≤ $\Delta G_{OH*}$ < 2.4 eV and $\Delta G_{O*}$ ≥ 3.52 eV) (**Figure 5d**) with the 12 catalysts meeting these standards (**Supplementary Dataset 4**). These catalysts were further analyzed using the microkinetic model (**Figure 6c**). All the selected catalysts exhibited 2e⁻ WOR activity and selectivity, with those marked in red having been experimentally validated for catalytic activity. $Sc_6O_{12}$ and $Y_8O_{18}$ showed the highest activity, occupying the top regions of both models, demonstrating their exceptional catalytic performance and selectivity. To validate these predictions, we used DFT to calculate the adsorption energies of these catalysts. **Figure 6d** displays the deviation distribution between predicted $\Delta G_{OH*}$ and $\Delta G_{O*}$ values and those obtained from DFT calculations (**Supplementary Note 12**). Remarkably, these deviations are within 0.2 eV, indicating a high correlation between the predicted and DFT-calculated energy distributions. This consistency confirms that our predictive model provides reasonable and accurate predictions of adsorption free energies for various systems.

Following the screening process, we compared the overpotentials of the top 16 screen-out catalysts with best predicted performance with those predicted in previous literature. Most of these catalysts (excluding alloys) are non-precious metal catalysts (**Supplementary Table 3**). Notable examples include $Sc_6O_{12}$, $Ni_{12}O_{18}$, Ni@Pd-$N_4$, $NaSnO_3$, and $Au_9Ag_3$, all of which exhibit $\eta_{WOR}$ overpotentials close to 0 V, with catalytic activity and selectivity comparable to the best-performing catalysts reported in the literature, such as ZnO(1010), $WO_3$, and $CaSnO_3$.[15,75,17] Although the materials reported in the literature exhibit excellent theoretical 2e$^-$ WOR activity, some display slightly higher overpotentials compared to the promising materials identified in our selection. These findings emphasize the effectiveness of our screening criteria, particularly in identifying non-precious metal materials with enhanced 2e$^-$ WOR performance.

We observed that the four materials are located at the top, highlighting their exceptional catalytic activity and selectivity (**Figure 6a**). This result not only reaffirms the well-established advantages of metal oxides and perovskites in 2e$^-$ WOR but also, for the first time, highlights the potential of SACs as effective 2e$^-$ WOR catalysts, which can provide new insights and references for future material exploration and experimental research. Furthermore, our research methodology, validated through multiple screening processes, confirms the reliability of the approach and the accuracy of the predictive model, demonstrating its substantial utility in identifying catalysts with application potential.

In summary, this screening process not only identifies potential 2e$^-$ WOR catalysts but also optimizes the selection of candidate materials through stringent standards, ensuring that the selected catalysts exhibit outstanding activity and selectivity in 2e$^-$ WOR reactions. This process offers reliable candidate materials for subsequent experimental studies and provides strong theoretical support for the efficient production of $H_2O_2$.

## Conclusion

In this study, we have presented a universal framework for catalyst design by developing a ML automation workflow with refined atomic environment descriptors (wACSF). As summarized in **Figure 1**, starting with the use of the *DigCat* database for extracting available experimental and computational data, the frameworks can use our proposed wACSF descriptors with advanced feature analyses and ML to universally capture the relationship between the atomic environment and the adsorption free energies of OH* and O*, across a wide range of materials including metal alloys, metal oxides and perovskites, and SACs. This framework can be used to directly predict the catalytic activities of 2e$^-$ WOR across different types of materials, together with our developed high-accuracy microkinetic volcano model that has passed the experimental benchmarking analysis in terms of the experimental exchange current density as a function of electrode potential and $\Delta G_{OH*}$. Following a series of strict criteria, high-performance 2e$^-$ WOR catalysts were predicted *via* a computational high-throughput process based on the input of ML-predicted $\Delta G_{OH*}$. Finally, DFT calculations were performed to validate the selected finalist candidates after the screening process.

All in all, this study has successfully developed a universal framework to realize an effective catalyst design for a wide range of materials at the same time, by integrating database application, universal local environment descriptors, ML modeling, microkinetic modeling, experimental benchmarking analysis, and computational high-throughput screening. This framework is general and proven effective for predicting $\Delta G_{OH*}$ and $\Delta G_{O*}$ across various materials, which is expected to be also applicable for the materials design of other catalysis including ORR and OER, where $\Delta G_{OH*}$ and $\Delta G_{O*}$ serve as the key descriptors.

## Data availability

All relevant data in this study are available in the Supplementary Materials and https://github.com/Weijie-Yang/FG-LAED.

## Code availability

The custom code developed in this study is available at https://github.com/Weijie-Yang/FG-LAED. The microkinetic models of 2e$^-$ WOR is available in our CatMath platform: https://catmath.cloud/. All the experimental data and computational structures can also be found in our DigCat database: https://www.digcat.org/.

# Acknowledgements

This work was funded by the Outstanding Youth Team Project Fund (No. 2752023YQ001), the Distinguished Youth Science Foundation of Hebei Province (No. E2023502015), the National Natural Science Foundation of China (Nos. 52176104 and 52006073), JSPS KAKENHI (Nos. JP23K13703, JP23K13599, JP24K23068, and JP24K23069), and the Hirose Foundation. The authors acknowledge the Center for Computational Materials Science, Institute for Materials Research, Tohoku University for the use of MASAMUNE-IMR (Nos. 202312-SCKXX-0203, 202312-SCKXX-0207, and 202312-SCKXX-0204) and the Institute for Solid State Physics (ISSP) at the University of Tokyo for the computational resources.

# Additional information

**Competing interests:** The authors declare no competing interests.